\pdfoutput=1
\documentclass{JINST}
\usepackage[subrefformat=parens]{subcaption}

\title{Extra-large crystal emulsion detectors for future large-scale experiments}

\author{
T. Ariga$^a$\thanks{Corresponding author.}~, 
A. Ariga$^a$,
K. Kuwabara$^b$, 
K. Morishima$^b$,
M. Moto$^b$, 
A. Nishio$^b$,
P. Scampoli$^{a,c}$,
M. Vladymyrov$^a$

\\
\llap{$^a$}Albert Einstein Center for Fundamental Physics, Laboratory for High Energy Physics, University of Bern, Sidlerstrasse 5, 3012 Bern, Switzerland\\
\llap{$^b$}F-lab, Nagoya University, 464-8602 Nagoya, Japan\\
\llap{$^c$}Department of Physics \lq{\lq{Ettore Pancini}\rq}\rq, University of Napoli Federico II, Complesso Universitario di Monte S. Angelo, 80126 Napoli, Italy\\

E-mail: \email{tomoko.ariga@lhep.unibe.ch}}

\abstract{Photographic emulsion is a particle tracking device which features the best spatial resolution among particle detectors. For certain applications, for example muon radiography, large-scale detectors are required. Therefore, a huge surface has to be analyzed by means of automated optical microscopes. An improvement of the readout speed is then a crucial point to make these applications possible and the availability of a new type of photographic emulsions featuring crystals of larger size is a way to pursue this program. This would allow a lower magnification for the microscopes, a consequent larger field of view resulting in a faster data analysis. In this framework, we developed new kinds of emulsion detectors with a crystal size of 600-1000 nm, namely 3-5 times larger than conventional ones, allowing a 25 times faster data readout. The new photographic emulsions have shown a sufficient sensitivity and a good signal to noise ratio. The proposed development opens the way to future large-scale applications of the technology, e.g. 3D imaging of glacier bedrocks or future neutrino experiments.}


\begin{document}

\section{Introduction}

Emulsion detectors feature the best spatial resolution among  particle detectors. 
They were extensively and successfully used in the past \cite{emul_review2,moire} and interesting present applications of the technology require large-scale detectors, that is emulsion areas of order 100-1000 m$^2$. This is the case, for example, of imaging of large objects by means of cosmic-ray muons (muon-radiography) or the detection of neutrinos. Muon radiography of volcanoes in Japan using emulsions was already performed \cite{asama2,okubo} and the first attempt to perform 3D imaging of glacier bedrocks in the Swiss Alps is currently under way \cite{SNSF}.
A crucial point for these new applications is the time needed to analyze a huge area of emulsions which is currently too slow for these applications even using automated scanning microscopes. Therefore, these new large-scale experiments require an increase of the speed performances for the automatic systems to become feasible.

The scanning speed ($s$), in general, can be described as follows, 
\[ s = freq\cdot FOV = \frac{1}{t_{image\ taking}+t_{stage\ movement}}\cdot( NpixelsX\cdot NpixelsY\cdot d^2) 
\]
The microscope data taking frequency ($freq$) is defined by the time of taking image ($t_{image\ taking}$) and mechanical stage movement ($t_{stage\ movement}$). The field of view ($FOV$) is defined by the multiplying the total number of pixels ($NpixelsX \cdot NpixelsY $) and the effective pixel area ($d^2$), here $d$ is a one-dimensional effective pixel size in specimen coordinate system.  Until now all the efforts to raise the readout speed were devoted to the improvement of the microscope scanning systems, namely $t_{image\ taking}$, $t_{stage\ movement}$ and $Npixels$ \cite{suts,ess,hts}. However, the remaining parameter, $d$ has been kept untouched.

Here we propose a complementary approach which is to increase the effective pixel size $d$ by enlarging the size of crystal of emulsion detectors. As the scanning speed is proportional to $d^2$, a considerable gain in speed is expected. Since we would like to observe a grain on more than one pixel of camera, the effective pixel size is usually set about a half of grain diameter. For example, for grain diameter of 500 nm after the development, about $d=250\ \textrm{nm/pixel}$ was used. The large grain detector would allow to choose a large $d$.
A possible drawback could be a degradation of angular resolution. The achieved angular resolution with a conventional film is 0.35 mrad \cite{ariga}, which is by far better than the resolution of about 5 mrad required for muon radiography. It will not be a problem to worsen it by a factor of 10.

In this framework, in 2014, a joint R\&D activity of the Laboratory for High Energy Physics (LHEP) of the University of Bern and the Fundamental Particle Physics Laboratory (F-lab) of the Nagoya University has started aiming at developing larger crystals up to 5 -10 times bigger than conventional ones.
The size of crystals used for the neutrino oscillation experiment OPERA was 200 nm \cite{opera_film} and was not larger than 300 nm for the previous experiments.
A recent study on slightly larger (350 - 400 nm) crystals gave useful suggestions on the crystal growth characteristics and showed that a good sensitivity can be reached~\cite{yokokura}.
The results reported in this paper will allow to increase the field of view by decreasing the magnification of the objective lenses. Here we report the first results on the characterization of emulsion particle detectors with a crystal size of 600-1000 nm which is 3-5 times larger than those of standard films.

\section{Production of emulsion gel with large silver halide crystals}

\begin{figure}[htbp]
\centering
\begin{minipage}[b]{0.38\linewidth}
\centering
\includegraphics[keepaspectratio, width=\textwidth]{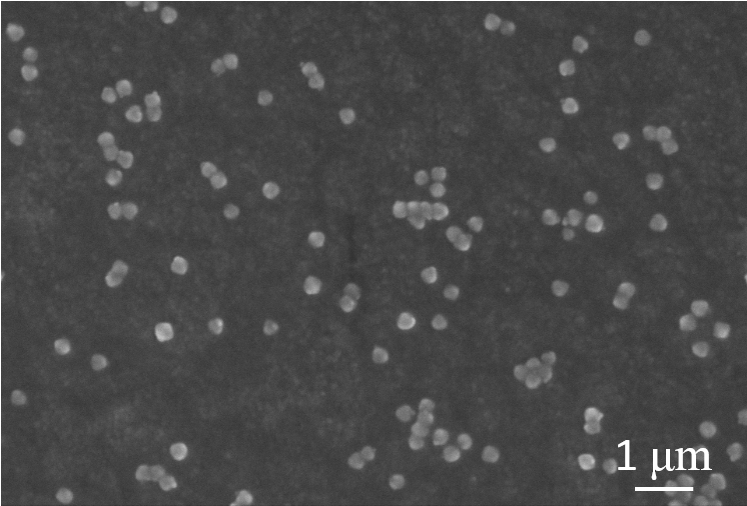}
\subcaption{Standard}\label{fig:sem-standard}
\end{minipage}
\begin{minipage}[b]{0.38\linewidth}
\centering
\includegraphics[keepaspectratio, width=\textwidth]{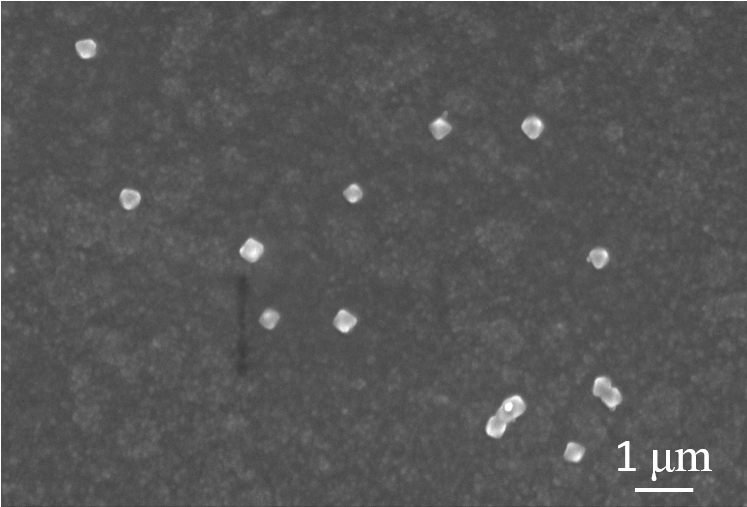}
\subcaption{Sample-1}\label{fig:sem-1}
\end{minipage}
\\
\begin{minipage}[b]{0.38\linewidth}
\centering
\includegraphics[keepaspectratio, width=\textwidth]{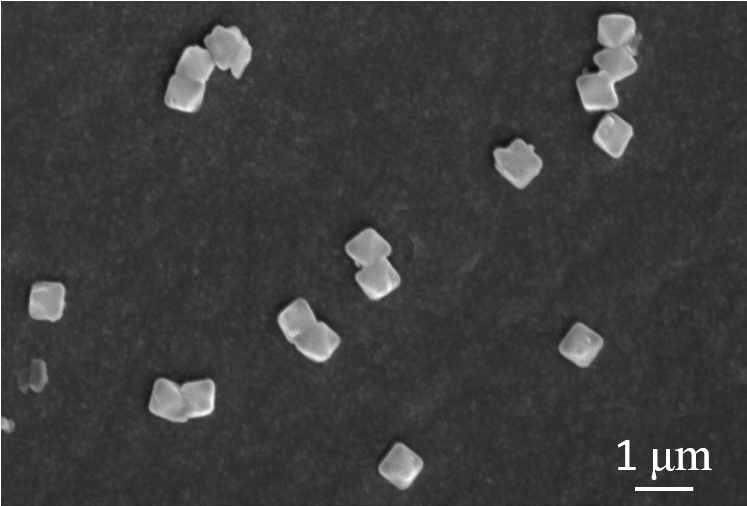}
\subcaption{Sample-2}\label{fig:sem-2}
\end{minipage}
\begin{minipage}[b]{0.38\linewidth}
\centering
\includegraphics[keepaspectratio, width=\textwidth]{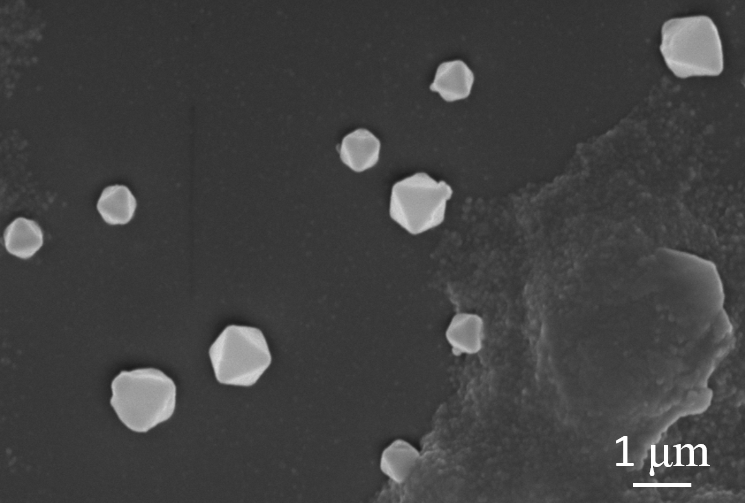}
\subcaption{Sample-3}\label{fig:sem-3}
\end{minipage}
\caption{
Electron microscope pictures of silver halide crystals in a standard gel (a), and in new samples after the three steps of the crystal growing procedure (b), (c) and (d). 
}\label{fig:sem}
\end{figure}

\begin{figure}[htbp]
\centering
\includegraphics[width=0.75\textwidth]{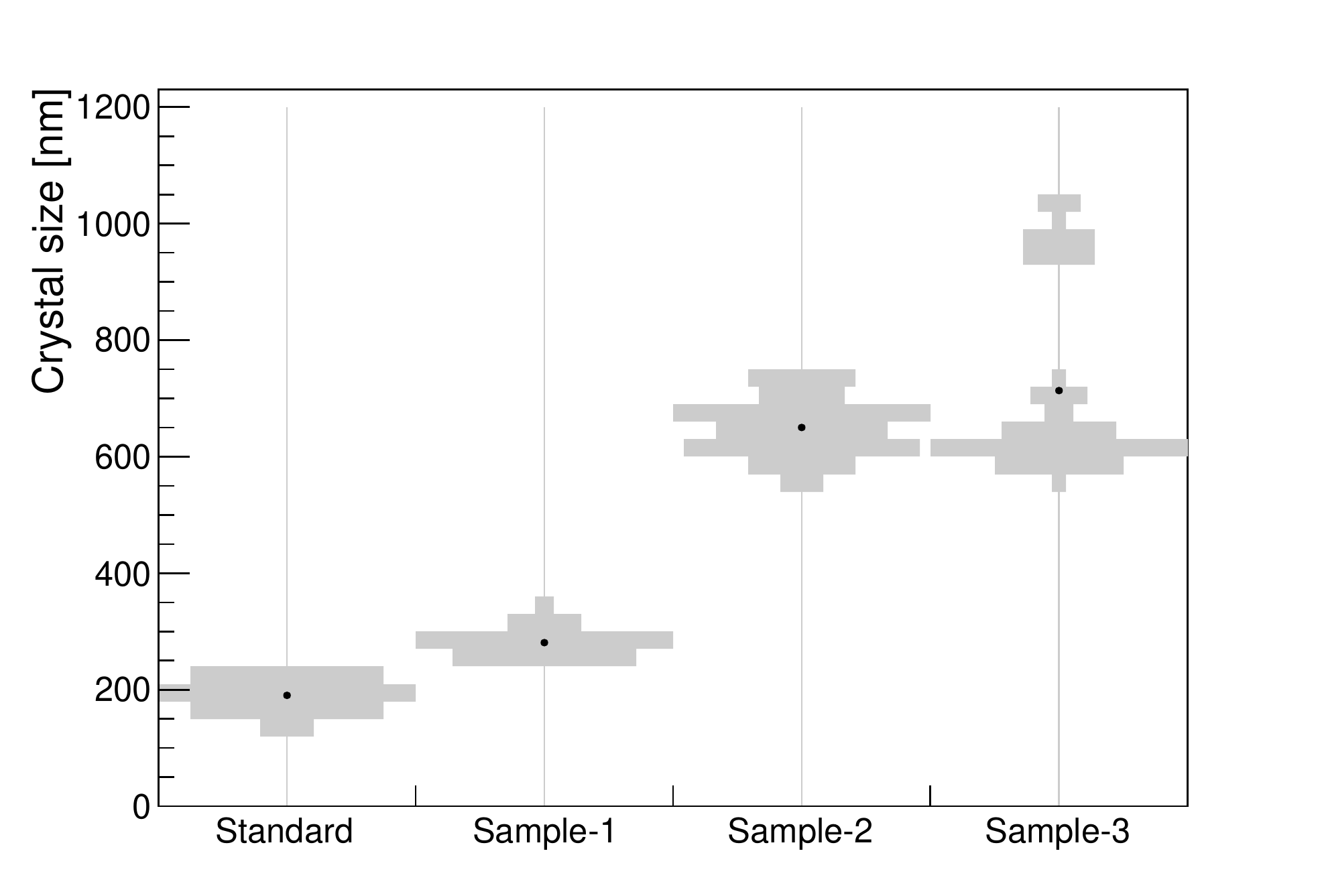}
\caption{
Distributions of the crystal diameters for a standard gel and the new emulsions (sample-1, 2, 3). The length of gray horizontal bar shows the population of entries, and the black dot shows the mean value of each sample.
} \label{fig:crystal_size}
\end{figure}

Emulsion detectors consist of silver halide (AgX) crystals homogeneously dispersed in a gelatin support, whose standard size is about 200 nm.  Each crystal acts as an independent charged particle detector and a particle track is made visible after the chemical development under an optical microscope in terms of metallic silver filaments (for a review on the technique, see \cite{emul_review2}). Briefly, as far as the emulsion production is concerned, a gel with silver halide crystals is the result of a double decomposition reaction between water-soluble silver salt and water-soluble halide salt to form insoluble AgX grains in an aqueous gelatin solution \cite{tani}.
The existing gel production machine is dedicated to the production of crystals with diameters below 250 nm. In order to increase the size of the AgX crystals, a multi-step procedure was followed, each step consisting of the above mentioned reactions involving a water-soluble silver salt and a water-soluble halide salt with a controlled double-jet precipitation method \cite{tani}. At the first stage, an emulsion gel with a crystal size of about 280 nm was produced (sample-1). Successively, a crystal growth was performed by using one eighth of the sample-1, which was kept un-sensitized, as a seed (sample-2). The additionally added silver and halide ions were deposited on the seed crystals and the crystal size became bigger. Finally, a further growth was done with one fourth of sample-2 (sample-3). During the process, the temperature of the gel, the speed of the solutions and the amount of the solvent are accurately kept under control and suitably adjusted.
The pictures of the AgX crystals taken by a scanning electron microscope are shown in Fig.~\ref{fig:sem} for each step. 
The area of each crystal was measured by an ad-hoc program and the distributions are reported in Fig.~\ref{fig:crystal_size} in terms of the diameter, assuming a circular shape for the crystals.
In particular, the mean size of sample-2  was (651$\pm$7)~nm, while sample-3 revealed two peaks in the diameter distribution with mean values of (628$\pm$6)~nm and (980$\pm$9)~nm. At the end of the production process some chemical products containing gold and sulfur were added in order to increase the sensitivity of the gel to minimum ionizing particles. This is the usual procedure but in the new samples the amount of chemicals was decreased by a factor of 6 with respect to conventional emulsions because of the reduced amount of crystals in the gel.

\section{Characterization of the new emulsions}

\begin{figure}[htbp]
\centering
\includegraphics[width=0.66\textwidth]{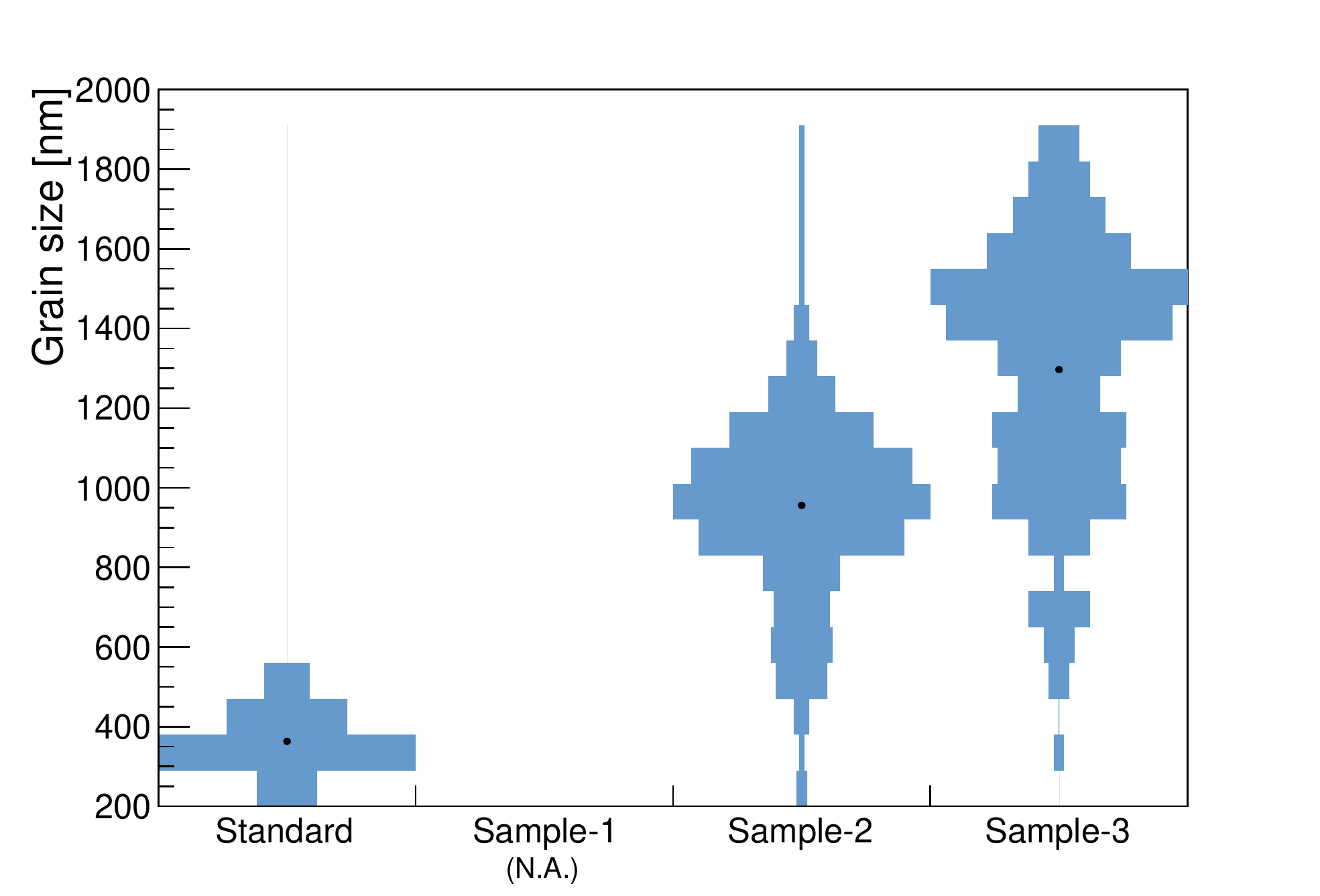}
\caption{
Distributions of the grain size in the developed films made of a standard gel, sample-2 and sample-3.
}\label{fig:grain_size}
\end{figure}

\begin{figure}[htbp]
\centering
\includegraphics[width=0.55\textwidth]{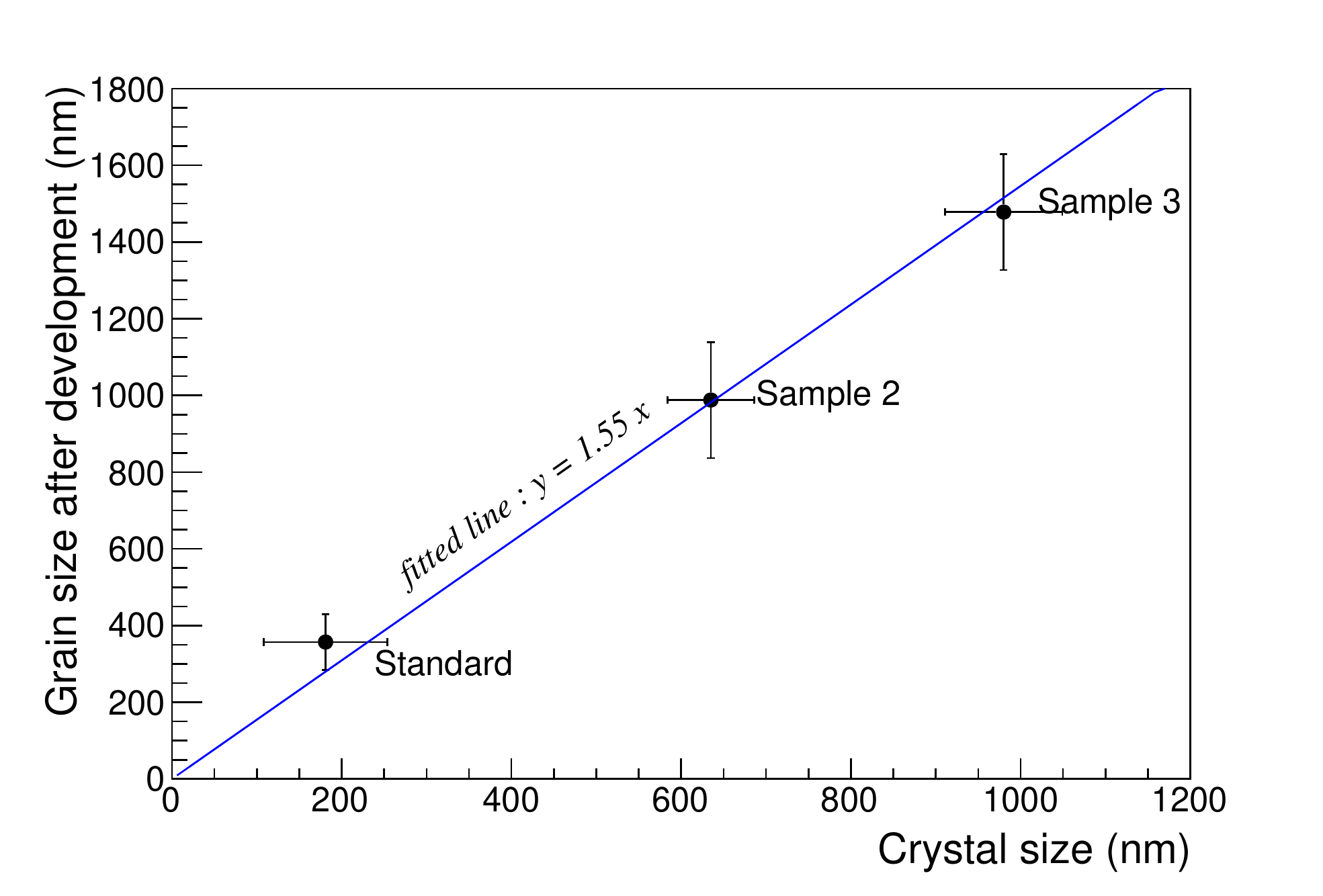}
\caption{
Relation between crystal size and grain size after the development. For sample 3, only the larger peak is used. 1$\sigma$ of the fitted Gaussians are used for the error bars.
}\label{fig:relation}
\end{figure}

The new emulsion samples were tested exposing the detectors to minimum ionizing particles at the UVSOR facility \cite{uvsor} in Japan, providing electrons with a mean energy of 100 MeV. Sample-1 was not exposed because it only was an intermediate stage for producing the samples with larger crystals. 
All samples were developed for 20 minutes with the OPERA developer by FUJI at a temperature of 20$\ {}^\circ\mathrm{C}$.
After the chemical development, the silver grain dimensions were measured with the same method described for Fig.~\ref{fig:crystal_size}. Their distributions are reported in Fig.~\ref{fig:grain_size}, showing that 3-5 times bigger grain size was achieved. Each peak in Fig.~\ref{fig:grain_size} is fitted with a gaussian function to obtain the mean value. The relation between crystal size and grain size after development is then plotted in the Fig.~\ref{fig:relation}. A linear relation is observed as expected.

\begin{figure}[htbp]
\centering
\begin{minipage}[b]{0.6\linewidth}
\centering
\includegraphics[keepaspectratio, width=\textwidth]{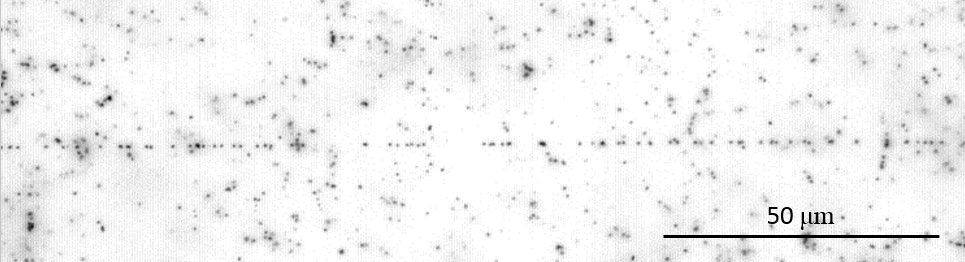}
\subcaption{Conventional film}\label{fig:trk_opera}
\end{minipage}
\\
\begin{minipage}[b]{0.6\linewidth}
\centering
\includegraphics[keepaspectratio, width=\textwidth]{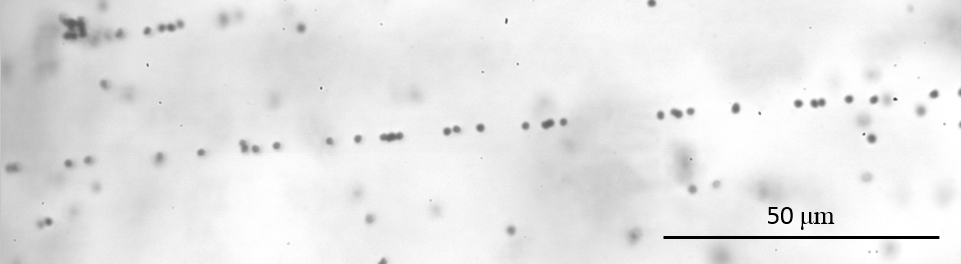}
\subcaption{Sample-2}\label{fig:trk_104}
\end{minipage}
\\
\begin{minipage}[b]{0.6\linewidth}
\centering
\includegraphics[keepaspectratio, width=\textwidth]{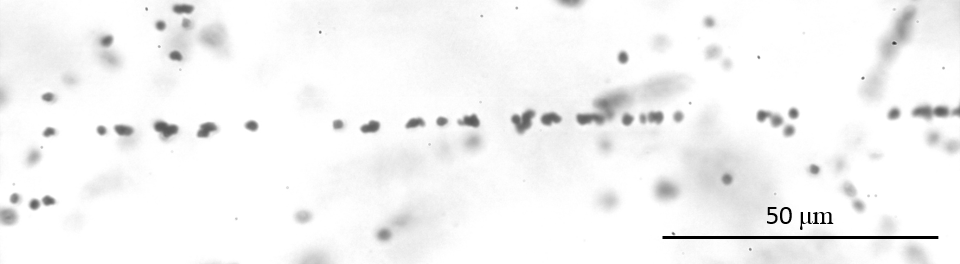}
\subcaption{Sample-3}\label{fig:trk_105}
\end{minipage}
\caption{
Electron tracks in a conventional film (a), and in the new samples (b) and (c).
}\label{fig:tracks}
\end{figure}

The performance of the new emulsion gels was then assessed in terms of sensitivity, background and intrinsic spatial resolution. In the first case the Grain Density (GD) was measured as the number of grains along 100 $\mu$m of a particle trajectory. In the second case, the Fog Density (FD) was measured as the number of the thermal induced grains per 1000 $\mu$m$^3$. Finally, the intrinsic spatial resolution was evaluated through the distribution of the distances between grains and a linear fit for a given track. The electron tracks for sample-2 and sample-3 are shown in Fig.~\ref{fig:tracks}, where an electron track in a conventional film \cite{opera_film} is also reported for comparison.

The GD and FD measured for the two samples are reported in Table 1. 
If crystal sensitivity does not depend on crystal size, the GD is expected to decrease with the number of crystals along the track, which is, in turn, inversely proportional to crystal size. In our case, the reduction of the GD is smaller than expected thanks to an improved crystal sensitivity.
Even though the sensitivity of the new emulsions is not as good as that showed by the conventional detector, 
previous studies \cite{fast} showed that in principle a track could be detected by using at least 5 grains. Assuming an emulsion layer thickness in the range (50-100) $\mu$m, the tracking efficiency for tracks with more than 10 grains should  be close to 100\%, then the sensitivity of the new samples is enough to clearly detect minimum ionizing particles. On the other side, an improved fog density was found with respect to standard films, assuring a good signal to noise ratio.

\begin{table}[hbtb]
\begin{center}
\begin{tabular}{|l|c|c|}
\hline
\  & GD (grains/100 $\mu$m) & FD (grains/1000 $\mu$m$^3$)\\
\hline
Conventional film & 36$\pm$1.1 & 2.9$\pm$0.3 \\
\hline
Sample-2 & 19$\pm$0.8 & 0.5$\pm$0.1 \\
\hline
Sample-3 & 20$\pm$0.8 & 0.6$\pm$0.1 \\
\hline
\end{tabular}
\caption{
Grain  and fog densities for the new films (sample-2 and sample-3) compared with a conventional film (\cite{opera_film}).
}\label{tb:gd_fd}
\end{center}
\end{table}

\begin{figure}[htbp]
\centering
\begin{minipage}[b]{0.49\linewidth}
\centering
\includegraphics[keepaspectratio, width=\textwidth]{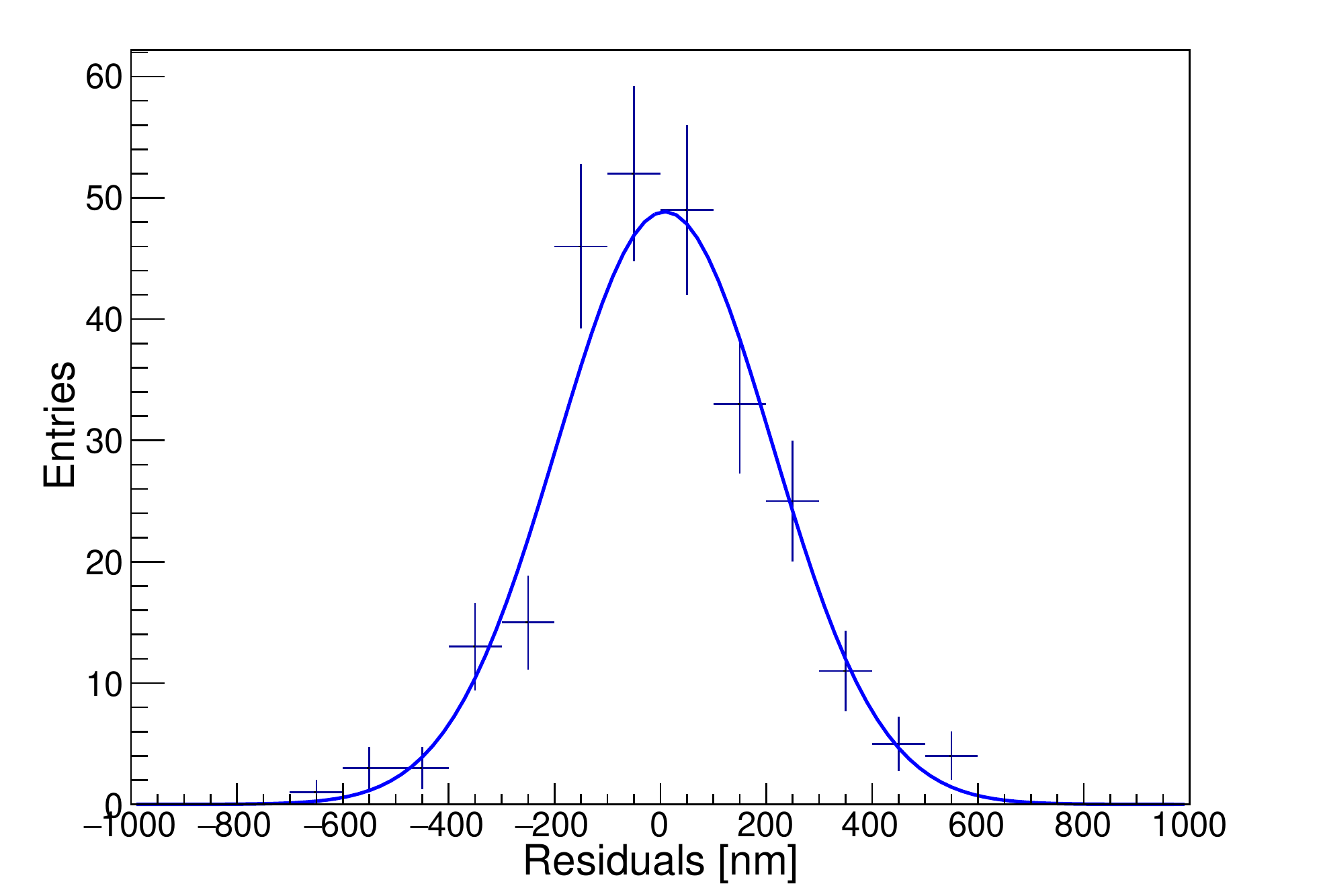}
\subcaption{Sample-2}\label{fig:resolution-sample2}
\end{minipage}
\begin{minipage}[b]{0.49\linewidth}
\centering
\includegraphics[keepaspectratio, width=\textwidth]{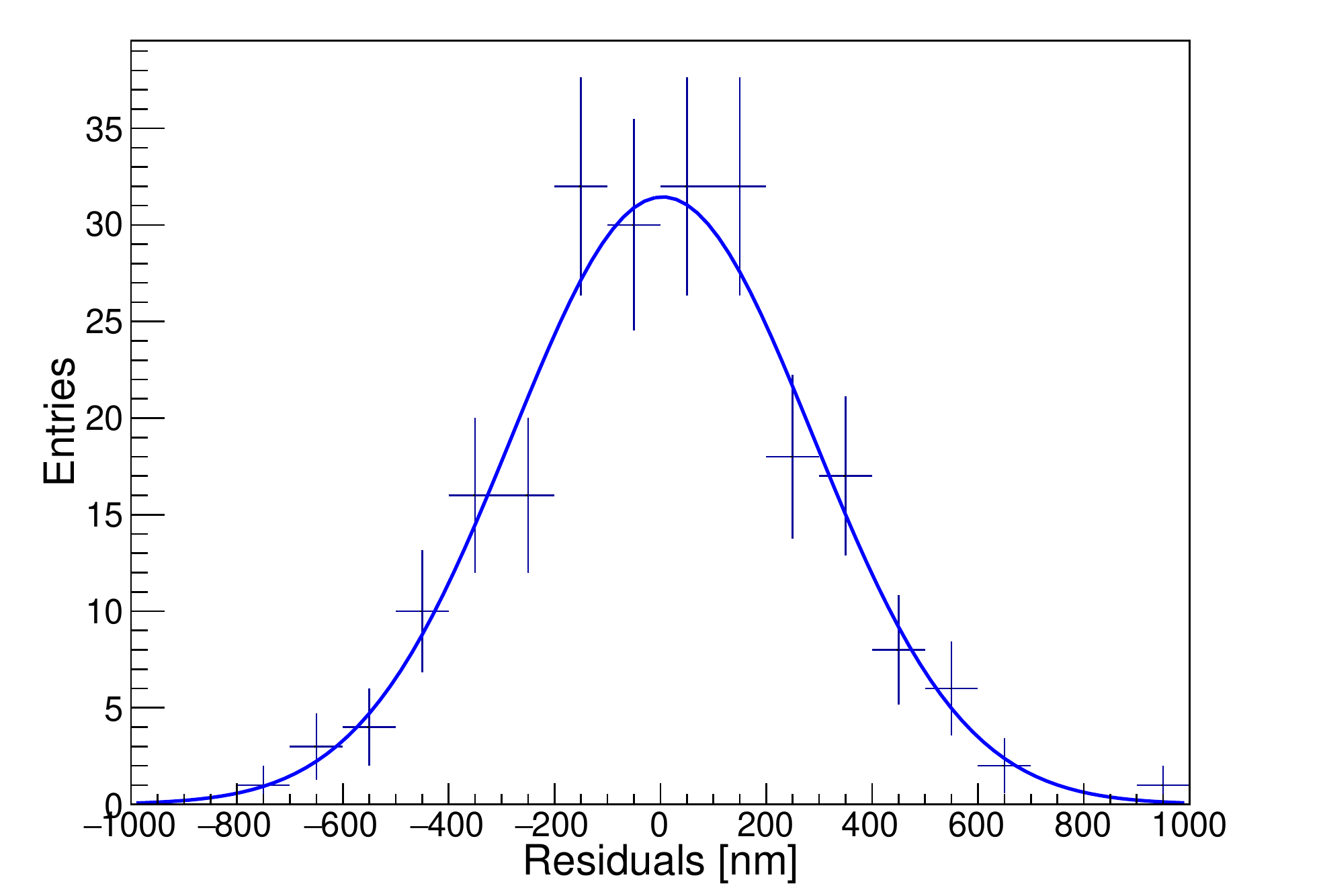}
\subcaption{Sample-3}\label{fig:resolution-sampl3}
\end{minipage}
\caption{
Deviation of grains from a linear fit of the electron tracks in sample-2 and sample-3.
}\label{fig:resolution}
\end{figure}

As far as the spatial resolution is concerned, conventional emulsions with 200 nm crystal diameter feature an intrinsic spatial resolution of 50-60 nm \cite{aegis_em, ariga}, which is expected to be worsened by the increased size of the silver grains. In Fig.~\ref{fig:resolution} we report the intrinsic spatial resolution for the two new samples. Values of about 200 nm and 280 nm where found for sample-2 and sample-3 respectively, showing that indeed the grain size deteriorates the resolution. Assuming a double-side coated emulsion film with a 400 $\mu$m-thick base, the angular resolution corresponds to 1 mrad for sample-3. Even so, it still remains suitable for the applications we are considering.

\section{Conclusions and future prospects}

We reported on the first production of emulsion detectors with AgX crystal sizes between 600 nm and 1000 nm, that is 3-5 times larger than those of  conventional emulsion detectors. Films with the new emulsion gels were characterized by exposing them to minimum ionizing particles (electrons with mean energy of 100 MeV), showing a sufficient sensitivity to minimum ionizing particles and a good signal to noise ratio. This new development is a key point for a very fast readout of particle tracks in the emulsion films. In particular, it allows a 25 times faster readout speed by using lower magnification objective lenses. 
Further studies on production method and long term stability are being performed. These new emulsion detectors open the way to future large-scale physics experiments and to their applications in geology such as the cosmic-ray muon radiography of the glacier bedrocks.

\acknowledgments
This work is supported by the Swiss National Science Foundation Ambizione grant PZ00P2\_154833.

\end{document}